\documentclass[runningheads]{llncs}

\usepackage{graphicx}
\usepackage{amsfonts}
\usepackage{amsmath}
\usepackage{amssymb}
\usepackage{enumerate}
\usepackage{bm}
\usepackage{color}
\usepackage{wrapfig}

\usepackage{hyperref}
\begin{document}

\title{Improved cross-validation for classifiers that make algorithmic choices to minimise runtime without compromising output correctness}  

%\title{Algorithmically generating new algebraic \\ features of polynomial systems \\ for machine learning
%\thanks{The authors are supported by EPSRC Project EP/R019622/1: \emph{Embedding Machine Learning within Quantifier Elimination Procedures}.}

%
\titlerunning{Improved cross-validation for classifiers that make algorithmic choices}

\author{Dorian Florescu \and Matthew England}
\authorrunning{D. Florescu and M. England}

\institute{Faculty of Engineering, Environment and Computing, \\Coventry University, Coventry, CV1 5FB, UK
\email{\\ \{Dorian.Florescu, Matthew.England\}@coventry.ac.uk}}

\maketitle              % typeset the header of the contribution

\begin{abstract}
Our topic is the use of machine learning to improve software by making choices which do not compromise the correctness of the output, but do affect the time taken to produce such output.  We are particularly concerned with computer algebra systems (CASs), and in particular, our experiments are for selecting the variable ordering to use when performing a cylindrical algebraic decomposition of $n$-dimensional real space with respect to the signs of a set of polynomials.

In our prior work we explored the different ML models that could be used, and how to identify suitable features of the input polynomials.  In the present paper we both repeat our prior experiments on problems which have more variables (and thus exponentially more possible orderings), and examine the metric which our ML classifiers targets.  The natural metric is computational runtime, with classifiers trained to pick the ordering which minimises this.  However, this leads to the situation were models do not distinguish between any of the non-optimal orderings, whose runtimes may still vary dramatically.  In this paper we investigate a modification to the cross-validation algorithms of the classifiers so that they do distinguish these cases, leading to improved results.  

%<250 words (currently 194)


\end{abstract}

\section{Introduction}
\label{SEC:Intro}

\subsection{Background and main thesis}

Machine Learning (ML), that is statistical techniques to give computer systems the ability to \emph{learn} rules from data, is a topic that has found great success in a diverse range of fields over recent years.
ML is most attractive when the underlying functional relationship to be modelled is complex or not well understood.  
With regards to the creation of software itself, while ML has a history of use for testing and security analysis \cite{GS17} it is less often used in the actual algorithms.  On the surface, this would be especially true for software that prizes mathematical correctness, such as computer algebra systems (CASs).  Here, a thorough understanding of the underlying relationships would seem to be a pre-requisite.  

However, CAS developers would acknowledge that their software actually comes with a range of options that, while having no effect on the correctness of the end result, can have a great effect on how long it takes to produce the result and exactly what form that result takes.  These choices range from the low level (in what order to perform a search that may terminate early) to the high (which of a set of competing exact algorithms to use for this problem instance).  

A well-known example is the choice of monomial ordering for a Gr\"{o}bner Basis.  This choice is actually quite abnormal in that there has been much study devoted to it and there are some clear pieces of advice to follow (e.g. that \texttt{degrevlex} ordering is the easiest to compute, and that if a \texttt{lex} ordering is needed it would be best to first compute a \texttt{degrevlex} basis and then convert).  A better example of the choices we consider would be the underlying variable order that is required to define any monomial ordering, for which there exists no such clear advice.  

In practice these less understood choices are usually either left entirely to the user, taken by human-made heuristics based on some experimentation (e.g. \cite{DSS04}), or made according to \emph{magic constants} where crossing a single threshold changes system behaviour \cite{Carette2004}.  Our main thesis is that many of these decisions could be improved by allowing ML algorithms to analyse the data.

\subsection{Outline of the paper and contribution}

Our experiments concern variable orderings for another prominent symbolic computation algorithm:  Cylindrical Algebraic Decomposition (CAD).  CAD is an expensive procedure, with the choice of ordering affecting not only computation time but often the tractability of even considering a problem.  We introduce the necessary background on CAD and its orderings in Section \ref{SEC:Background}.  
We describe our prior work using ML to make this choice \cite{HEWDPB14}, \cite{HEWBDP19}, \cite{EF19}, \cite{FE19a} in Section \ref{SEC:Prior} which includes experimenting with a range of ML models, and developing techniques to generate suitable features from the input data.  This prior work was all conducted on a large dataset of 3-variable problems (a choice from 6 orderings).

The new contributions of the present paper are two-fold.  First, we have applied our prior methodology to a dataset of 4-variable problems (choice from 24 orderings) and we report on how it handled this increased complexity.  Secondly, we examine and improve the training goal of our ML classifiers.  
The natural metric for this problem is runtime, and our old classifiers are trained to pick the ordering which minimises this for a given CAD input.  However, this meant our training did not distinguish between any of the non-optimal orderings even though the difference between these could be huge.  In Section \ref{SEC:TimeCV} we report on a new cross-validation approach for our classifiers which aims to make them aware of these \emph{different shades of wrong} and thus make choices which reduce the overall runtime even if the number of problems where the classifiers pick the absolute best runtime is unchanged.   

In Section \ref{SEC:MLMethodology} and \ref{SEC:Results} we describe the methodology and results respectively for our new experiments on choosing the variable ordering for 4-variable CAD problems, both with and without the new cross-validation approach.  We also compare against the best known human-made heuristics.

\section{Background on variable ordering for CAD}
\label{SEC:Background}

\subsection{Cylindrical algebraic decomposition}
\label{SUBSEC:CAD}

A \emph{Cylindrical Algebraic Decomposition} (CAD) is a \emph{decomposition} of ordered $\mathbb{R}^n$ space into cells arranged \emph{cylindrically}: the projections of any pair of cells with respect to the variable ordering are either equal or disjoint.  I.e. the projections all lie within cylinders over the cells of an induced CAD of the lower dimensional space.  All these cells are (semi)-algebraic meaning each can be described with a finite sequence of polynomial constraints.  

A CAD is usually produced to be \emph{truth-invariant} for a logical formula, meaning the formula is either true or false on each cell.  Such a decomposition can then be used to analyse the formula, and for example, perform Quantifier Elimination (QE) over the reals.  I.e. given a quantified Tarski formula in prenex normal form we can find an equivalent quantifier free formula over the reals by building a CAD for the quantifier-free part of the formula, querying a finite number of sample points (one from each cell), and then using the corresponding cell descriptions.  For example, QE could transform $\exists x, ax^2 + b x + c = 0 \land a \neq 0$ to the equivalent unquantified statement $b^2 - 4ac \geq 0$ by building a CAD of $(x,a,b,c)$.  In practice, the quantifier free equivalent would come as the conjunction of several parts (one from each cell) which logically simplify to the stated result.  

CAD was introduced by Collins in 1975 \cite{Collins1975} and works relative to a set of polynomials.  Collins' CAD produces a decomposition so that each polynomial has constant sign on each cell (thus truth-invariant for any formula built with those polynomials).  The algorithm first projects the polynomials into smaller and smaller dimensions; and then uses these to lift $-$ to incrementally build decompositions of larger and larger spaces according to the polynomials at that level.  There have been a great many developments in the theory and implementation of CAD since Collins' original work which we do not describe here.  The collection \cite{CJ98} summarises the work up to the mid-90s while the second author's journal articles \cite{BDEMW16} \cite{EBD19} attempt summaries of CAD progress since in their introduction and background sections.  
CAD is the backbone of all QE implementations as it is the only implemented complete procedure for the problem.  QE has numerous applications throughout science and engineering\footnote{Recently even economics too \cite{MBDET18}, \cite{MDE18}.}
\cite{Sturm2006} which would in turn benefit from faster CAD.  Our work also speeds up independent applications of CAD, such as reasoning with multi-valued functions \cite{DBEW12}, motion planning \cite{WDEB13}, and identifying multistationarity in biological networks \cite{BDEEGGHKRSW17}, \cite{BDEEGGHKRSW20}.

\subsection{Variable ordering}
\label{SUBSEC:VarOrd}

The definition of cylindricity and both stages of the CAD algorithm are relative to an ordering of the variables.  
For example, given polynomials in variables ordered as $x_n \succ x_{n-1} \succ \dots, \succ x_2 \succ x_1$ we first project away $x_n$ and so on until we are left with polynomials univariate in $x_1$.  
We then start lifting by decomposing the $x_1-$axis, and then the $(x_1, x_2)-$plane and so so on.  The cylindricity condition refers to projections of cells in $\mathbb{R}^n$ onto a space $(x_1, \dots, x_m$) where $m<n$. 
As noted above there have been numerous advances to CAD since its inception but the need for a fixed variable ordering remains.

Depending on the application, the variable ordering may be determined, constrained, or free. QE, requires that quantified variables are eliminated first and that variables are eliminated in the order in which they are quantified.  However, variables in blocks of the same quantifier (and the free variables) can be swapped, so there is partial freedom. In the example discussed in Section \ref{SUBSEC:CAD} we may use any variable ordering that projects the quantified variable $x$ first to perform the QE and discover the discriminant.  A CAD for the quadratic polynomial under ordering $a \prec b \prec c$ has only 27 cells, but we need 115 for the reverse ordering.

This choice of variable ordering can have a great effect on the time and memory use of CAD, and the number of cells in the output (how course or fine the decomposition is).  In fact, Brown and Davenport presented a class of problems in which one variable ordering gave output of double exponential complexity in the number of variables and another output of a constant size \cite{BD07}.  

Heuristics have been developed to choose a variable ordering, with Dolzmann \textit{et al.} \cite{DSS04} giving the best known study.  After analysing a variety of metrics they proposed a heuristic, \texttt{sotd}, which constructs the full set of projection polynomials for each permitted ordering and selects the ordering whose corresponding set has the lowest \textbf{s}um \textbf{o}f \textbf{t}otal \textbf{d}egrees for each of the monomials in each of the polynomials.   The second author demonstrated examples for which that heuristic could be misled in \cite{BDEW13}; and then later showed that tailoring to an implementation could improve performance \cite{EBDW14}.  These heuristics all involved potentially costly projection operations on the input polynomials.

Another human-made heuristic was proposed by Brown in his ISSAC 2004 tutorial notes \cite{Brown2004}.  This  chooses a variable ordering according to the following criteria, starting with the first and breaking ties with successive ones.
\begin{enumerate}[1)]
\item Eliminate a variable first if it appears with the lowest overall individual degree in the input.
\item For each variable calculate the maximum total degree (i.e. sum of the individual degrees) for the set of terms in the input 
in which it occurs.  Eliminate first the variable for which this is lowest.
\item Eliminate a variable first if there is a smaller number of terms in the input  which contain the variable.
\end{enumerate}

The Brown heuristic is far cheaper than the \texttt{sotd} heuristic (because the latter performs projections before measuring degrees).  Surprisingly, our experiments on CAD problems in 3-variables all suggest that the Brown heuristic makes better choices than \texttt{sotd} (even before one considers the time taken to run the heuristic itself). This counter-intuitive finding does not generalise into our 4-variable problem set, as discussed later.

\section{Prior ML work on this problem}
\label{SEC:Prior}

\subsection{Results from CICM 2014}

The first application of ML for choosing a CAD variable ordering was \cite{HEWDPB14} which used a support vector machine to select which of three human-made heuristics to follow.  
The SVM considered 11 simple algebraic features of the input polynomials (mostly different measures of degree and variable occurrence).  The experiments were on 3-variable CAD problems and although the Brown heuristic was found to make the best choices on average, the experiments identified substantial subsets of examples for which each of the three heuristics outperformed the others.  The key conclusion was that the machine learned choice did significantly better than any one heuristic overall.

\subsection{Results from CICM 2019}

The present authors revisited these experiments earlier this year in \cite{EF19}.  We used the same dataset but this time ML was used to predict directly the variable ordering for CAD, rather than choosing a heuristic.  The motivation for picking a heuristic in \cite{HEWDPB14} was that if the methodology were applied to problems with more variables it would still mean making a choice from 3 possibilities rather than an exponentially growing number.  However, upon investigation there were many problems where none of the human-made heuristics made good choices and so savings could be made by considering all possible orderings\footnote{Of course, this methodology will have to be changed to deal with higher numbers of variables but since CAD is rarely tractable with more than 5 variables this is not a particularly pressing concern.  We note that there are several meta-algorithms that may be applicable to sample the possible ordering without evaluating them all.  For example, a Monte Carlo tree searched was used in \cite{KUV15} to sample the possible multivariate Horner schemes and pick an optimal one in the CAS FORM.}.

In \cite{EF19} we also considered a more diverse selection of ML methods than \cite{HEWDPB14}.  We experimented with four common ML classifiers: 
%\begin{itemize}
%\item 
K$-$Nearest Neighbours (KNN);
%\item 
Multi-Layer Perceptron (MLP);
%\item 
Decision Tree (DT); and
%\item 
Support Vector Machine (SVM) with RBF kernel, all using the same set of 11 features from \cite{HEWDPB14}.

The results showed that all three of the new models performed better substantially better than the SVM (the only classifier to be tried before); and that all four classifiers outperformed the human-made heuristics.

\subsection{Results from SC-Square 2019}

We next considered how to extract further information from the input data.  The 11 features used in \cite{HEWDPB14}, \cite{EF19} were inspired by Brown's heuristic \cite{Brown2004} (e.g. measures of variable degree and frequency of occurrence).  In particular, they can all be cheaply extracted from polynomials.  

In \cite{FE19a} a new feature generation procedure was presented, based on the observation that the original features can be formalised mathematically using a small number of basic functions (average, sign, maximum) evaluated on the degrees of the variables in either one polynomial or the whole system.  Considering all possible combinations of these functions led to 78 useful and independent features for our 3-variable dataset.  The experiments were repeated with these, with the results showing that all four ML classifiers improved their predictions.  

Using these new features the choices of the best performing classifier allowed CAD to solve all problems in the testing set with a runtime only 6\% more than the best possible (i.e. the time taken if the optimal ordering were used for every problem).  Using only the original features, the choices of the best ML classifier led to 14\% more than the minimum runtime.  Following the choices of Brown's heuristic led to runtimes 27\% more than the minimum. 

\subsection{Related work on ML for mathematical software}

The work described above is the only published work on ML for choosing a CAD variable ordering.  There are only a handful of other examples of ML within CASs:  \cite{HEDP16}, \cite{HEWBDP19} on the question of whether to precondition CAD with Gr\"{o}bner Bases; \cite{KIMA16} on deciding the order of sub-formulae solving for a QE procedure; and \cite{KUV15} on choosing a multivariate Horner scheme.  Other areas of mathematical software have made more use of ML.  For example, in the mathematical logic community the ML-selected portfolio SAT solver \textsc{SATZilla} \cite{XHHL08} is well-known, while more recently \textsc{MapleSAT} views solver branching as an optimisation problem to be tackled with ML \cite{LHPCG17}.  There are also several examples of ML within the automated reasoning community (see e.g. \cite{Urban2007},  \cite{KBKU13}, \cite{BHP14}).  A survey on ML for mathematical software was presented at ICMS 2018 \cite{England2018}.

\section{New cross-validation based on computing times}
\label{SEC:TimeCV}

\subsection{Motivation}
\label{SUBSEC:CVmotivate}

In all of the authors previous ML experiments for CAD \cite{HEWDPB14}, \cite{EF19}, \cite{FE19a}, the models were optimised simply to predict which of the possible variable orderings leads to the smallest computing time for CAD.  This is not an ideal approach:
\begin{itemize}
\item First, runtimes for CAD, like all software, will contain a degree of noise from various hardware and software factors.  While it is common for a given CAD problem to have a wide range of possible runtimes depending on the ordering, that does not mean that all orderings give runtimes distinct from the others.  The runtimes commonly appear in clusters.  Thus it is often the case that the smallest runtime be only slightly lower than the second smallest, and that difference could well be down to noise.  Thus when training to target only the very quickest runtime we risk exaggerating the effects of such noise.

\item Second, during training, when a model makes an incorrect prediction this could mean selecting an ordering that produces a runtime very close to the optimal or another that is significantly larger.  The training would not distinguish between these cases $-$ there is no distinction between picking an ``almost good'' ordering and a ``very bad'' ordering.  However, from the point of view of a user judging these selections there is a big difference!
\end{itemize}

One of the traditional metrics used to evaluate an ML classifier is \textit{accuracy}, defined as the number of test examples for which the classifier makes the correct choice.  In our context, correct means picking the optimal variable ordering from the $n!$ possibilities.  We recognised that for our application this definition of accuracy is not sufficient to judge the classifiers and so in our prior work we also presented the total CAD runtime for the testing set when using the variable orderings of a classifier (which we referenced in the summary above).  

The anonymous referees of our earlier papers commented that perhaps accuracy could be redefined into something more appropriate for our application.  For example, judge a classifier as being correct for a problem instance if it picks an ordering which produces a runtime within $x$\% of the minimum runtime that can be achieved for that instance\footnote{In Section \ref{SEC:MLMethodology} we use $x=20$ but we are still debating the most appropriate value.}.   This led us to consider whether the training algorithms could be adapted to take account of this more nuanced definition of accuracy.  We decided to introduce this in the stage of the methodology where cross-validation is used for hyperparameter selection: a single technique that is used for all of the different ML classifiers we work with.

\subsection{Traditional ML cross-validation}
\label{SUBSEC:TradCV}

We describe first the typical procedure of cross-validation used when preparing a ML classifier which sets the parameters and hyperparameters of a model.

The parameters are variables that can be fine-tuned so that the prediction error reaches a local or global minimum in the parameter space. For example,  the weights in an artificial neural network or the support vectors in an SVM.
The hyperparameters are model configurations selected before training. They are often specified by the practitioner based on experience or a heuristic,  e.g. the number of layers in a neural network or the value of $k$ in a $k$-nearest neighbour model. The connection between the hyperparameters and the model prediction is more complex, and thus, typically, these are tuned using grid search in the hyperparameter space to minimise the prediction error. 

To prevent the situation where the model returns poor results on new datasets not used in training, also known as overfitting, the hyperparameters and parameters are tuned on different datasets. The typical approach is cross-validation.

In $G$-fold cross-validation (see for example the introduction of \cite{Bishop2006}), the data is split into $G$ groups of equal size $M$:
\begin{align}
\mathcal{D}_1 &= \left\lbrace \boldsymbol{f}^{(k_1^1)},\dots,\boldsymbol{f}^{(k_1^M)} \right\rbrace \nonumber \\
&\hspace*{0.075in} \vdots \label{EQ:groups} \\
\mathcal{D}_G &= \left\lbrace \boldsymbol{f}^{(k_G^1)},\dots,\boldsymbol{f}^{(k_G^M)} \right\rbrace, \nonumber
\end{align}
where each group entry is a vector of features for a problem instance:
\[
\boldsymbol{f}^{(k_g^m)} = \left[f_1^{(k_g^m)},\dots,f_{n_f}^{(k_g^m)}\right], 
\qquad g = 1, \dots, G, \quad m = 1, \dots, M.
\]
Each entry in such a vector is a scalar number and $n_f$ denotes the number of features we derive for each instance.  See \cite{FE19a} for details of the features we use and how they are generated from the polynomials.  

Let $c^{(k_g^m)}$ denote the target class corresponding to data point $\boldsymbol{f}^{(k_g^m)}$.
%, for $m=1,\dots,M, g=1,\dots,G.$
An ML classifier with parameters $ \boldsymbol{\theta} $ is modelled as a function $\mathcal{M}^{\boldsymbol{\theta}}: $ $\mathbb{R}^{n_f}\rightarrow \{1,2,\dots,n_c\}$, where $n_c$ denotes the number of classes. In our context the number of classes is the number of CAD variable orderings acceptable for the underlying application.  

Typically, the classifier also depends on a number of hyperparameters that can each take a finite number of values. Here, we will denote by $H$ the number of all possible hyperparameter combinations, such that $\mathcal{M}^{\boldsymbol{\theta}}_h$, $h=1,\dots,H,$ denotes the classifier with parameters $\boldsymbol{\theta}$ and hyperparameters defined by index $h$.
The typical cross-validation procedure trains the parameters $\boldsymbol{\theta}$ of classifiers $\left\lbrace\mathcal{M}^{\boldsymbol{\theta}}_h\right\rbrace_{h=1}^H$ on each combination of $G-1$ data groups in \eqref{EQ:groups}, adding up to $G \cdot H$ models.

Let $\hat{c}_h^{(k_g^m)}$ denote $\mathcal{M}^{\boldsymbol{\theta}_g}_h\left( \boldsymbol{f}^{(k_g^m)} \right)$, the class prediction of a classifier whose parameters were trained on the dataset $\mathcal{D}_1\cup\dots\cup\mathcal{D}_{g-1}\cup \mathcal{D}_{g+1}\cup\dots\cup\mathcal{D}_G$. Then the optimal $h_{opt}$ is computed by maximising the following quantity:
\begin{equation}
\label{EQ:reg_score}
h_{opt}=\underset{h}{\text{argmax}}\left(\frac{1}{G}\sum_{g=1}^G \text{score}_h^g\right),
\end{equation}
where $\text{score}_h^g=\text{score}\left(\hat{c}_h^{(k_g^m)},c^{(k_g^m)}\right)$, and $ \text{score}(\cdot,\cdot) $ denotes the F1-score of group $ G $ for the model prediction \cite{Chinchor1992}.
In other words, the typical cross-validation procedure identifies the hyperparameters that maximise the performance of the model at predicting the very best ordering. I.e. it does not take into account the actual computing time of the prediction $-$ just whether it was the quickest. 

\subsection{Adapted ML cross-validation}
\label{SUBSEC:NewCV}

Our change to the cross-validation procedure is to instead calculate $h_{opt}$ as
\begin{equation}
\label{EQ:cust_score}
h_{opt}=\underset{h}{\text{argmax}}\left(\frac{1}{G}\sum_{g=1}^G -\text{ctime}_h^g\right),
\end{equation}
where $\text{ctime}_h^g=\frac{1}{M}\sum_m \text{ctime}\left(k_g^m,\hat{c}^{(k_g^m)}\right)$, and $\text{ctime}\left(k_g^m,\hat{c}^{(k_g^m)}\right)$ denotes the recorded time for computing CAD on data point $\boldsymbol{f}^{(k_g^m)}$ using the variable ordering given by class prediction $\hat{c}^{(k_g^m)}$. By evaluating the computing time for all data points, this cross-validation method penalises the variable orderings leading to very large computing times, but does not penalise the ones close to the optimum.  Thus we do not expect the change to affect how often a classifier chooses the optimal ordering, but it should improve the choices made in cases where the optimum is missed.

\section{ML experiment methodology}
\label{SEC:MLMethodology}

We describe a ML experiment to choose the variable ordering for CAD.  The methodology used is similar to that of our recent paper \cite{FE19a} except that (a) we use a dataset of 4-variable problems instead of 3-variable ones; and (b) we ran the classifiers with both the original and the adapted cross-validation procedure.

\subsection{Problem set}
\label{SUBSEC:dataset}

We are working with the \texttt{nlsat} dataset\footnote{Freely available from \url{http://cs.nyu.edu/~dejan/nonlinear/}} produced to evaluate the work in \cite{JdM12}, thus the problems are all fully existentially quantified.  
Although there are CAD algorithms that reduce what is being computed based on the quantification structure (e.g. Partial CAD \cite{CH91}), the conclusions we draw are likely to generalise.  

We selected the $2080$ problems with $4$ variables, meaning each has a choice of $24$ different variable orderings.  We extracted only the polynomials involved, and randomly divided into two datasets for training ($ 1546 $) and testing ($ 534 $).  Only the former is used to tune the ML model parameters and hyperparameters.

\subsection{Software}
\label{SUBSEC:software}

We work with the CAD routine \texttt{CylindricalAlgebraicDecompose}: part of the \texttt{RegularChains} Library for \textsc{Maple}.   It builds decompositions first of $\mathbb{C}^n$ before refining to a CAD of $\mathbb{R}^n$ \cite{CMXY09}, \cite{CM14b}, \cite{BCDEMW14}.  We ran the code in Maple $ 2018 $ but used an updated version of the \texttt{RegularChains} Library (\url{http://www.regularchains.org}).  
Brown's heuristic and the features for ML were coded in the \texttt{sympy} package 
%\cite{SymPy2017} 
v1.3 for Python 2.7.  The \texttt{sotd} heuristic was implemented in \textsc{Maple} as part of the \texttt{ProjectionCAD} package \cite{EWBD14}. 
Training and evaluation of the ML models was done using the  \texttt{scikit-learn} package \cite{SciKitLearn2011} v0.20.2 for Python 2.7.  In order to implement our adapted cross-validation procedure we had to rewrite a number of the standard commands within the package to both use the redefined $h_{opt}$ in (\ref{EQ:cust_score}), and to access the data it requires during the cross-validation.

\subsection{Timings}
\label{SUBSEC:Timing}

Each individual CAD was constructed by a Maple script called separately from Python (to avoid any Maple caching of results).  The target variable ordering for ML was defined as the one that minimises the computing time for a given problem.  
All CAD function calls included a time limit. For the training dataset an initial time limit of $16$ s was used, which was doubled if all orderings timed out (a target variable ordering could be assigned for all problems using time limits no bigger than $32$ s).
The problems in the testing dataset were all processed with a single larger time limit of $ 64 $ s for all orderings, with any problems that timed out having their runtime recorded as $64$s. 

\subsection{Computing the features}
\label{SUBSEC:FeatureSimp}

We computed algorithmically the set of features for $4$ variables $\{f^{(i)} \}_{i=1}^{n_f}$ where $n_f=1440$, using the procedure introduced in \cite{FE19a}.

Given the set of problems $\{\boldsymbol{Pr}_1,\dots,\boldsymbol{Pr}_N\}, N=1546, $ some of the features $f^{(i)}$ turn out to be constant, i.e. 
$
f^{(i)}(\boldsymbol{Pr}_1)=f^{(i)}(\boldsymbol{Pr}_2)=\dots=f^{(i)}(\boldsymbol{Pr}_N).
$
Such features will have no benefit for ML and are removed.  Further, other features may be repetitive, i.e.
$
f^{(i)}(\boldsymbol{Pr}_n)=f^{(j)}(\boldsymbol{Pr}_n),\forall n =1,\dots,N,
$  and are merged into one single feature.
%This repetition may represent a mathematical equality, or just be the case of the given dataset.  Either way, they are merged into a single feature for the experiment. 
After this step, we are left with $105$ features.

\subsection{ML models}

We choose commonly used deterministic ML models for this experiment (for details on the methods see e.g. the textbook \cite{Bishop2006}).  
%We also give an overview of each in \cite{EF19}.
\begin{itemize}
\item The K$-$Nearest Neighbours (KNN) classifier \cite[\S 2.5]{Bishop2006}.
\item The Decision Tree (DT) classifier \cite[\S 14.4]{Bishop2006}.
\item The Multi-Layer Perceptron (MLP) classifier \cite[\S 2.5]{Bishop2006}.
\item The Support Vector Machine (SVM) classifier with Radial Basis Function (RBF) kernel \cite[\S 6.3]{Bishop2006}. 
\end{itemize}
We fixed the RBF kernel for SVM as it was found to produce better results than other basis functions for a similar problem of learning from algebraic feastures in \cite{Bridge2010}, and including basis choice in cross-validation creates a much larger search space.

Each model was trained using grid search $3$-fold cross-validation, i.e. the set was randomly divided into $3$ and each possible combination of $2$ parts was used to tune the model parameters, leaving the last part for fitting the hyperparameters with cross-validation, by optimising the average F-score.
Grid searches were performed for an initially large range for each hyperparameter; then gradually decreased to home into optimal values.
The optimal hyperparameters selected during cross-validation are in Table \ref{tab:hyper}.

\begin{table}[t]
	\caption{The ML hyperparameters optimised on the training dataset using the standard cross-validation (CV) routine and the new CV routine.  \label{tab:hyper}}
\begin{tabular}{|c|c|c|c|}
\hline
\textbf{Model} & \textbf{Hyperparameter} & \textbf{Value} (standard CV) & \textbf{Value} (new CV) \\
\hline
\hline
Decision Tree 	& Criterion 		 & Entropy & Gini impurity	\\			
				& Maximum tree depth & $ 6 $		 & $14$\\
\hline
K-Nearest  & Train instances weighting & Inversely proportional & Inversely proportional\\		
Neighbours &                           & to distance            & to distance \\
	& Algorithm & Ball Tree & Ball Tree\\
	& Number of neighbours & $13$ & $14$\\
\hline
SVM  & Regularization para. $C$ & $2.41$ & $1.66$\\		
	& Basis para. $\gamma$ & $0.0097$ & $0.0097$\\
\hline
Multi-Layer  & Hidden layer size & $18$ & $17$\\		
Perceptron	& Activation function & Hyperbolic Tangent & Identity\\
	& Regularization para. $\alpha$ & $ 1\cdot10^{-4} $ & $ 1\cdot10^{-4} $\\
\hline
		\end{tabular}	
\end{table}

\subsection{Evaluating the ML models and human-made heuristics}
\label{SUBSEC:Human}

The ML models will be compared on two metrics: \textit{Accuracy}, defined as the percentage of problems where a model's predicted variable ordering led to a computing time closer than $20\%$ of the time it took the optimal ordering; and \textit{Time} defined as the total time taken to evaluate all problems in the test set using that model's predictions for variable ordering.  We note that Accuracy is defined differently in our prior work \cite{EF19}, \cite{FE19a} where we measured only how often a heuristic picked the very best ordering.  

We will also test the two best-known human constructed heuristics \cite{Brown2004}, \cite{DSS04} described in Section \ref{SUBSEC:VarOrd}. Unlike the ML models, these can end up predicting several variable orderings (when they cannot discriminate).  In practice if this were to happen the heuristic would select one randomly (or perhaps lexicographically), however that final pick is not meaningful.   To accommodate this we evaluate these heuristics as follows:
\begin{itemize}
\item For each problem, the prediction accuracy of such a heuristic is judged to be the the percentage of its predicted variable orderings that are also target orderings (i.e. within 20\% of the minimum).  The average of this percentage over all problems in the testing dataset represents the prediction accuracy. 
\item Similarly, the computing time for such methods is assessed as the average computing time over all predicted orderings, and it is this that is summed up for all problems in the testing dataset.
\end{itemize}

\section{Experimental Results}
\label{SEC:Results}

The results are presented in Table \ref{tab:results}.  Each ML model appears twice in the top table via its acronym with each of the following appended:
\begin{description}
\item[$-$O:] for one trained with the original (and typical) ML cross-validation method based on \eqref{EQ:reg_score} as was used in our prior work \cite{EF19}, \cite{FE19a}.
\item[$-$N:] for one trained by the new cross-validation approach described in Section \ref{SUBSEC:NewCV} which is based on computing time as in \eqref{EQ:cust_score}.
\end{description}

The bottom table details the two human-constructed heuristics along with the outcome of a random choice between the $24$ orderings.  
%Of course, these do not change with the form of cross-validation.
We might expect a random choice to be correct once in $24$ times of the time but it is higher as for some problems there were multiple variable orderings with equally fast timings.

The minimum total computing time, achieved if we select an optimal ordering for every problem, is $2,177$s.  This is what would be achieved by the \emph{Virtual Best Heuristic}. Choosing at random would take $ 8,291 $s, almost 4 times as much. The maximum time, if we selected the worst ordering for every problem (the \emph{Virtual Worst Heuristic}), is $22,735$s. The Decision Tree model trained with the new cross validation achieved the shortest time of all with $3,627$s, $ 67\% $ more than the minimal possible. 

The recorded time taken by each model to make a prediction, which is included in the timings reported in Table \ref{tab:results}, varied greatly between ML and the heuristics. The prediction time for the heuristics was $286$ s for \texttt{sotd} and $23$ s for \texttt{Brown}. In contrast, the total time taken by the ML to make predictions was less than one second for all models.

\begin{table}[b]
	\caption{Performance on the testing dataset of the ML classifiers (using both the standard and new cross-validation routines), the human-made heuristics, and a random choice.  The virtual best and worst solvers show the range of possibilities. \label{tab:results}}

	\begin{center}
		\begin{tabular}{|l||c|c||c|c||c|c||c|c|}
			\hline
	& \,DT-O\, & \,DT-N\, & KNN-O & KNN-N 
	& MLP-O & MLP-N & SVM-O & SVM-N 
	\\
			\hline
	\textbf{Accuracy} 
	& $51.7\%$ & $54.3\%$ & $53.9\%$ & $54.5\%$ 
	& $53.6\%$ & $56.9\%$ & $53.9\%$ & $54.9\%$ 
	\\
	\textbf{Time (s)} 
	& $4,022$  & $3,627$  & $3,808$  & $3,748$  
	& $3,972$   & $3,784$   & $3,795$   & $3,672$  
	\\				
			\hline
		\end{tabular}
	\end{center}

	\begin{center}
		\begin{tabular}{|c|c|c|c|c|c|}
			\hline
			 &  Virtual Best & Virtual Worst & \,random\, &  \,\texttt{Brown}\, & \,\texttt{sotd}\, \\
			\hline
	\textbf{Accuracy}  & $100\%$ & $0\%$    & $17.0\%$ & $20.1\%$ & $47.8\%$ \\
	\textbf{Time (s)}  & $2,177$ & $22,735$ & $8,291$  & $8,292$  & $4,348$   \\				
			\hline
		\end{tabular}
	\end{center}
	
\end{table}

\subsection{Results of new cross-validation method}

For each ML model the performance when trained with the new cross-validation was better (measured using either of our metrics) than when trained with the original procedure.  The scale of the improvement varied:  the timings of the decision tree reduced by 9.8\% but those of the KNN classifier only by 1.6\%. 

Thus we can conclude the new methodology to be beneficial.  However, we note that it is still the case that our two metrics do not agree on the best model:  DT-N achieved the lowest times but KNN-N the highest accuracy.  The latter is better at picking a good (within 20\% of the minimal) ordering but when it fails to do so it makes mistakes of greater magnitude.  So there is scope for further work to make our ML models take into account the full range of possibilities.  It may be that this requires a tailored approach to the training of parameters in each different classifier.

\subsection{Comparison of Brown and sotd on the 4-variable dataset}
 
Of the two human-made heuristics, \texttt{Brown} performed far worse than \texttt{sotd}.  This is the opposite of the findings in \cite{HEWDPB14}, \cite{EF19}, \cite{FE19a} for 3-variable problems. This is not necessarily in conflict: the added information taken by \texttt{sotd} will grow in size exponentially with the variables, and thus we would expect the predictive information it carries to be more valuable. However, the cost of \texttt{sotd} will also be increasing rapidly denting this value.  The time taken by \texttt{sotd} to make all the predictions is $286\ s$, while the time for \texttt{Brown} is less than $10\%$ of that at $23\ s$.  For this dataset at least, it is well worth paying the price of \texttt{sotd} as the savings over Brown's heuristic are far more substantial. 

\subsection{Value of ML on the 4-variable dataset}

All heuristics (ML and human-made) are further away from the optimum on this 4-variable dataset than they were on the three variable one, to be expected given we are choosing from 24 rather than 6 orderings.  Our best performing model achieves timings $67\%$ greater than the minimum (it was 6\% for 3-variable problems).  However, the best human-made heuristic had timings 98\% greater.  

In fact, every ML model outperformed both the human constructed heuristics in regards to both metrics, and when using either the original or the new cross-validation approach.  So we can easily conclude that our ML methodology generalises to 4-variable problems.  However, it also clear that there is much more scope for future improvement.

\section{Summary}
\label{SEC:Conc}

We have demonstrated that our methodology of ML for choosing a CAD variable ordering may be applied to 4-variable problems where it continues its dominance over human-made heuristics.  We have also presented an addition to the ML training methodology to better reflect our application domain and demonstrated the benefit of this experimentally.  This new methodology could be applied to any ML application which seeks to make a choice to minimise computational runtime.

\subsubsection*{Acknowledgements}  

This work is funded by EPSRC Project EP/R019622/1: \emph{Embedding Machine Learning within Quantifier Elimination Procedures}.  %We thank the anonymous referees for their comments which improved the paper.


\bibliographystyle{splncs04}
\bibliography{CAD}

\begin{thebibliography}{10}
\providecommand{\url}[1]{\texttt{#1}}
\providecommand{\urlprefix}{URL }
\providecommand{\doi}[1]{https://doi.org/#1}

\bibitem{Bishop2006}
Bishop, C.: Pattern Recognition and Machine Learning. Springer (2006)

\bibitem{BCDEMW14}
Bradford, R., Chen, C., Davenport, J., England, M., {Moreno~Maza}, M., Wilson,
  D.: Truth table invariant cylindrical algebraic decomposition by regular
  chains. In: Gerdt, V., Koepf, W., Seiler, W., Vorozhtsov, E. (eds.) Computer
  Algebra in Scientific Computing, Lecture Notes in Computer Science,
  vol.~8660, pp. 44--58. Springer International Publishing (2014),
  \url{http://dx.doi.org/10.1007/978-3-319-10515-4_4}

\bibitem{BDEEGGHKRSW17}
Bradford, R., Davenport, J., England, M., Errami, H., Gerdt, V., Grigoriev, D.,
  Hoyt, C., Ko\v{s}ta, M., Radulescu, O., Sturm, T., Weber, A.: A case study on
  the parametric occurrence of multiple steady states. In: Proceedings of the
  2017 ACM International Symposium on Symbolic and Algebraic Computation. pp.
  45--52. ISSAC '17, ACM (2017), \url{https://doi.org/10.1145/3087604.3087622}

\bibitem{BDEEGGHKRSW20}
Bradford, R., Davenport, J., England, M., Errami, H., Gerdt, V., Grigoriev, D.,
  Hoyt, C., Ko\v{s}ta, M., Radulescu, O., Sturm, T., Weber, A.: Identifying the
  parametric occurrence of multiple steady states for some biological networks.
  Journal of Symbolic Computation  \textbf{98},  84--119 (2020),
  \url{https://doi.org/10.1016/j.jsc.2019.07.008}

\bibitem{BDEMW16}
Bradford, R., Davenport, J., England, M., McCallum, S., Wilson, D.: Truth table
  invariant cylindrical algebraic decomposition. Journal of Symbolic
  Computation  \textbf{76},  1--35 (2016),
  \url{http://dx.doi.org/10.1016/j.jsc.2015.11.002}

\bibitem{BDEW13}
Bradford, R., Davenport, J., England, M., Wilson, D.: Optimising problem
  formulations for cylindrical algebraic decomposition. In: Carette, J.,
  Aspinall, D., Lange, C., Sojka, P., Windsteiger, W. (eds.) Intelligent
  Computer Mathematics, Lecture Notes in Computer Science, vol.~7961, pp.
  19--34. Springer Berlin Heidelberg (2013),
  \url{http://dx.doi.org/10.1007/978-3-642-39320-4_2}

\bibitem{Bridge2010}
Bridge, J.: Machine learning and automated theorem proving. Tech. Rep.
  UCAM-CL-TR-792, University of Cambridge, Computer Laboratory (2010)

\bibitem{BHP14}
Bridge, J., Holden, S., Paulson, L.: Machine learning for first-order theorem
  proving. Journal of Automated Reasoning  \textbf{53},  141--172 (2014),
  \url{https://doi.org/10.1007/s10817-014-9301-5}

\bibitem{Brown2004}
Brown, C.: Companion to the tutorial: {C}ylindrical algebraic decomposition,
  presented at {ISSAC} '04. URL
  \url{http://www.usna.edu/Users/cs/wcbrown/research/ISSAC04/handout.pdf}
  (2004)

\bibitem{BD07}
Brown, C., Davenport, J.: The complexity of quantifier elimination and
  cylindrical algebraic decomposition. In: Proceedings of the 2007
  {I}nternational {S}ymposium on {S}ymbolic and {A}lgebraic {C}omputation. pp.
  54--60. ISSAC '07, ACM (2007), \url{https://doi.org/10.1145/1277548.1277557}

\bibitem{Carette2004}
Carette, J.: Understanding expression simplification. In: Proceedings of the
  2004 {I}nternational {S}ymposium on {S}ymbolic and {A}lgebraic {C}omputation.
  pp. 72--79. ISSAC '04, ACM (2004),
  \url{https://doi.org/10.1145/1005285.1005298}

\bibitem{CJ98}
Caviness, B., Johnson, J.: Quantifier Elimination and Cylindrical Algebraic
  Decomposition. Texts \& Monographs in Symbolic Computation, Springer-Verlag
  (1998), \url{https://doi.org/10.1007/978-3-7091-9459-1}

\bibitem{CM14b}
Chen, C., {Moreno Maza}, M.: An incremental algorithm for computing cylindrical
  algebraic decompositions. In: Feng, R., Lee, W., Sato, Y. (eds.) Computer
  Mathematics, pp. 199---221. Springer Berlin Heidelberg (2014),
  \url{https://doi.org/10.1007/978-3-662-43799-5_17}

\bibitem{CMXY09}
Chen, C., {Moreno Maza}, M., Xia, B., Yang, L.: Computing cylindrical algebraic
  decomposition via triangular decomposition. In: Proceedings of the 2009
  {I}nternational {S}ymposium on {S}ymbolic and {A}lgebraic {C}omputation. pp.
  95--102. ISSAC '09, ACM (2009), \url{https://doi.org/10.1145/1576702.1576718}

\bibitem{Chinchor1992}
Chinchor, N.: {MUC}-4 evaluation metrics. In: Proceedings of the 4th conference
  on Message Understanding ({MUC4 '92}). pp. 22--29. Association for
  Computational Linguistics. (1992),
  \url{https://doi.org/10.3115/1072064.1072067}

\bibitem{Collins1975}
Collins, G.: Quantifier elimination for real closed fields by cylindrical
  algebraic decomposition. In: Proceedings of the 2nd GI Conference on Automata
  Theory and Formal Languages. pp. 134--183. Springer-Verlag (reprinted in the
  collection \cite{CJ98}) (1975),
  \url{https://doi.org/10.1007/3-540-07407-4_17}

\bibitem{CH91}
Collins, G., Hong, H.: Partial cylindrical algebraic decomposition for
  quantifier elimination. Journal of Symbolic Computation  \textbf{12},
  299--328 (1991), \url{https://doi.org/10.1016/S0747-7171(08)80152-6}

\bibitem{DBEW12}
Davenport, J., Bradford, R., England, M., Wilson, D.: Program verification in
  the presence of complex numbers, functions with branch cuts etc. In: 14th
  International Symposium on Symbolic and Numeric Algorithms for Scientific
  Computing. pp. 83--88. SYNASC '12, IEEE (2012),
  \url{http://dx.doi.org/10.1109/SYNASC.2012.68}

\bibitem{DSS04}
Dolzmann, A., Seidl, A., Sturm, T.: Efficient projection orders for {CAD}. In:
  Proceedings of the 2004 {I}nternational {S}ymposium on {S}ymbolic and
  {A}lgebraic {C}omputation. pp. 111--118. ISSAC '04, ACM (2004),
  \url{https://doi.org/10.1145/1005285.1005303}

\bibitem{England2018}
England, M.: Machine learning for mathematical software. In: Davenport, J.,
  Kauers, M., Labahn, G., Urban, J. (eds.) Mathematical Software -- Proc. ICMS
  2018. Lecture Notes in Computer Science, vol. 10931, pp. 165--174. Springer
  International Publishing (2018),
  \url{https://doi.org/10.1007/978-3-319-96418-8_20}

\bibitem{EBD19}
England, M., Bradford, R., Davenport, J.: Cylindrical algebraic decomposition
  with equational constraints. Journal of Symbolic Computation
  \textbf{Accepted (In Press)} (2019),
  \url{https://doi.org/10.1016/j.jsc.2019.07.019}

\bibitem{EBDW14}
England, M., Bradford, R., Davenport, J., Wilson, D.: Choosing a variable
  ordering for truth-table invariant cylindrical algebraic decomposition by
  incremental triangular decomposition. In: Hong, H., Yap, C. (eds.)
  Mathematical Software -- ICMS 2014. Lecture Notes in Computer Science,
  vol.~8592, pp. 450--457. Springer Heidelberg (2014),
  \url{http://dx.doi.org/10.1007/978-3-662-44199-2_68}

\bibitem{EF19}
England, M., Florescu, D.: Comparing machine learning models to choose the
  variable ordering for cylindrical algebraic decomposition. In: Kaliszyk, C.,
  Brady, E., Kohlhase, A., Sacerdoti, C. (eds.) Intelligent Computer
  Mathematics. Lecture Notes in Computer Science, vol. 11617, pp. 93--108.
  Springer International Publishing (2019),
  \url{https://doi.org/10.1007/978-3-030-23250-4_7}

\bibitem{EWBD14}
England, M., Wilson, D., Bradford, R., Davenport, J.: Using the {R}egular
  {C}hains {L}ibrary to build cylindrical algebraic decompositions by
  projecting and lifting. In: Hong, H., Yap, C. (eds.) Mathematical Software --
  ICMS 2014. Lecture Notes in Computer Science, vol.~8592, pp. 458--465.
  Springer Heidelberg (2014),
  \url{http://dx.doi.org/10.1007/978-3-662-44199-2_69}

\bibitem{FE19a}
Florescu, D., England, M.: Algorithmically generating new algebraic features of
  polynomial systems for machine learning. In: Abbott, J., Griggio, A. (eds.)
  Proceedings of the 4th Workshop on Satisfiability Checking and Symbolic
  Computation ($\mathsf{SC}^2$ 2019). No.~2460 in CEUR Workshop Proceedings
  (2019), \url{http://ceur-ws.org/Vol-2460/}

\bibitem{GS17}
Ghaffarian, S., Shahriari, H.: Software vulnerability analysis and discovery
  using machine-learning and data-mining techniques: {A} survey. ACM Comput.
  Surv.  \textbf{50}(4) (2017), \url{https://doi.org/10.1145/3092566}

\bibitem{HEDP16}
Huang, Z., England, M., Davenport, J., Paulson, L.: Using machine learning to
  decide when to precondition cylindrical algebraic decomposition with
  {G}roebner bases. In: 18th International Symposium on Symbolic and Numeric
  Algorithms for Scientific Computing (SYNASC '16). pp. 45--52. IEEE (2016),
  \url{https://doi.org/10.1109/SYNASC.2016.020}

\bibitem{HEWBDP19}
Huang, Z., England, M., Wilson, D., Bridge, J., Davenport, J., Paulson, L.:
  Using machine learning to improve cylindrical algebraic decomposition.
  Mathematics in Computer Science  \textbf{13}(4),  461--488 (2019),
  \url{https://doi.org/10.1007/s11786-019-00394-8}

\bibitem{HEWDPB14}
Huang, Z., England, M., Wilson, D., Davenport, J., Paulson, L., Bridge, J.:
  Applying machine learning to the problem of choosing a heuristic to select
  the variable ordering for cylindrical algebraic decomposition. In: Watt, S.,
  Davenport, J., Sexton, A., Sojka, P., Urban, J. (eds.) Intelligent Computer
  Mathematics, Lecture Notes in Artificial Intelligence, vol.~8543, pp.
  92--107. Springer International (2014),
  \url{http://dx.doi.org/10.1007/978-3-319-08434-3_8}

\bibitem{JdM12}
Jovanovic, D., de~Moura, L.: Solving non-linear arithmetic. In: Gramlich, B.,
  Miller, D., Sattler, U. (eds.) Automated Reasoning: 6th International Joint
  Conference ({IJCAR}), Lecture Notes in Computer Science, vol.~7364, pp.
  339--354. Springer (2012), \url{https://doi.org/10.1007/978-3-642-31365-3_27}

\bibitem{KIMA16}
Kobayashi, M., Iwane, H., Matsuzaki, T., Anai, H.: Efficient subformula orders
  for real quantifier elimination of non-prenex formulas. In: Kotsireas, S.,
  Rump, M., Yap, K. (eds.) Mathematical Aspects of Computer and Information
  Sciences (MACIS '15). Lecture Notes in Computer Science, vol.~9582, pp.
  236--251. Springer International Publishing (2016),
  \url{https://doi.org/10.1007/978-3-319-32859-1_21}

\bibitem{KBKU13}
K{\"u}hlwein, D., Blanchette, J., Kaliszyk, C., Urban, J.: {MaSh}: {M}achine
  learning for sledgehammer. In: Blazy, S., {Paulin-Mohring}, C., Pichardie, D.
  (eds.) Interactive Theorem Proving, Lecture Notes in Computer Science,
  vol.~7998, pp. 35--50. Springer Berlin Heidelberg (2013),
  \url{https://doi.org/10.1007/978-3-642-39634-2_6}

\bibitem{KUV15}
Kuipers, J., Ueda, T., Vermaseren, J.: Code optimization in {FORM}. Computer
  Physics Communications  \textbf{189},  1--19 (2015),
  \url{https://doi.org/10.1016/j.cpc.2014.08.008}

\bibitem{LHPCG17}
Liang, J., {Hari Govind}, V., Poupart, P., Czarnecki, K., Ganesh, V.: An
  empirical study of branching heuristics through the lens of global learning
  rate. In: Gaspers, S., Walsh, T. (eds.) Theory and Applications of
  Satisfiability Testing -- {SAT} 2017, Lecture Notes in Computer Science, vol.
  10491, pp. 119--135. Springer International Publishing (2017),
  \url{https://doi.org/10.1007/978-3-319-66263-3_8}

\bibitem{MBDET18}
Mulligan, C., Bradford, R., Davenport, J., England, M., Tonks, Z.: Non-linear
  real arithmetic benchmarks derived from automated reasoning in economics. In:
  Bigatti, A., Brain, M. (eds.) Proceedings of the 3rd Workshop on
  Satisfiability Checking and Symbolic Computation ($\mathsf{SC}^2$ 2018). pp.
  48--60. No.~2189 in CEUR Workshop Proceedings (2018),
  \url{http://ceur-ws.org/Vol-2189/}

\bibitem{MDE18}
Mulligan, C., Davenport, J., England, M.: {T}heory{G}uru: {A} {M}athematica
  package to apply quantifier elimination technology to economics. In:
  Davenport, J., Kauers, M., Labahn, G., Urban, J. (eds.) Mathematical Software
  -- Proc. ICMS 2018. Lecture Notes in Computer Science, vol. 10931, pp.
  369--378. Springer International Publishing (2018),
  \url{https://doi.org/10.1007/978-3-319-96418-8_44}

\bibitem{SciKitLearn2011}
Pedregosa, F., Varoquaux, G., Gramfort, A., Michel, V., Thirion, B., Grisel,
  O., Blondel, M., Prettenhofer, P., Weiss, R., Dubourg, V., Vanderplas, J.,
  Passos, A., Cournapeau, D., Brucher, M., Perrot, M., Duchesnay, E.:
  Scikit-learn: {M}achine learning in {P}ython. Journal of Machine Learning
  Research  \textbf{12},  2825--2830 (2011),
  \url{http://www.jmlr.org/papers/v12/pedregosa11a.html}

\bibitem{Sturm2006}
Sturm, T.: New domains for applied quantifier elimination. In: Ganzha, V.,
  Mayr, E., Vorozhtsov, E. (eds.) Computer Algebra in Scientific Computing,
  Lecture Notes in Computer Science, vol.~4194, pp. 295--301. Springer Berlin
  Heidelberg (2006), \url{https://doi.org/10.1007/11870814_25}

\bibitem{Urban2007}
Urban, J.: {MaLARea: A} metasystem for automated reasoning in large theories.
  In: Empirically Successful Automated Reasoning in Large Theories (ESARLT
  '07), CEUR Workshop Proceedings, vol.~257, p.~14. CEUR-WS (2007),
  \url{http://ceur-ws.org/Vol-257/}

\bibitem{WDEB13}
Wilson, D., Davenport, J., England, M., Bradford, R.: A ``piano movers''
  problem reformulated. In: 15th International Symposium on Symbolic and
  Numeric Algorithms for Scientific Computing. pp. 53--60. SYNASC '13, IEEE
  (2013), \url{http://dx.doi.org/10.1109/SYNASC.2013.14}

\bibitem{XHHL08}
Xu, L., Hutter, F., Hoos, H., {Leyton-Brown}, K.: {SAT}zilla: {P}ortfolio-based
  algorithm selection for {SAT}. Journal Of Artificial Intelligence Research
  \textbf{32},  565--606 (2008), \url{https://doi.org/10.1613/jair.2490}

\end{thebibliography}

\end{document}